\pdfoutput=1
\documentclass[%
twocolumn,%
 amssymb, amsmath,%
 nobibnotes,
 aps,prl,%
 floatfix,
superscriptaddress
]{revtex4-1}

\usepackage{docs}%
\usepackage{bm}%
\usepackage[colorlinks=true,linkcolor=blue,citecolor=blue]{hyperref}%
\usepackage{graphicx}
\expandafter\ifx\csname package@font\endcsname\relax\else
 \expandafter\expandafter
 \expandafter\usepackage
 \expandafter\expandafter
 \expandafter{\csname package@font\endcsname}%
\fi
\begin{document}

\title{Uncompensated Polarization in Incommensurate Modulations of Perovskite Antiferroelectrics}%
\author{Tao Ma}%
\thanks{These authors contributed equally to this work.}
\affiliation{Ames Laboratory, U.S. Department of Energy, Ames, IA 50011, USA}%
\author{Zhongming Fan}
\thanks{These authors contributed equally to this work.}
\affiliation{Department of Materials Science and Engineering, Iowa State University, Ames 50011, Iowa, USA}
\author{Bin Xu}
\affiliation{Physics Department and Institute for Nanoscience and Engineering, University of Arkansas, Fayetteville, Arkansas 72701, USA}
\affiliation{School of Physical Science and Technology, Soochow University, Suzhou, Jiangsu 215006, China}
\author{Tae-Hoon Kim}
\affiliation{Ames Laboratory, U.S. Department of Energy, Ames, IA 50011, USA}%
\author{Ping Lu}
\affiliation{Sandia National Laboratories, PO Box 5800, MS 1411, Albuquerque, NM 87185, USA}
\author{Laurent Bellaiche}
\affiliation{Physics Department and Institute for Nanoscience and Engineering, University of Arkansas, Fayetteville, Arkansas 72701, USA}
\author{Matthew J. Kramer}
\affiliation{Ames Laboratory, U.S. Department of Energy, Ames, IA 50011, USA}
\affiliation{Department of Materials Science and Engineering, Iowa State University, Ames 50011, Iowa, USA}
\author{Xiaoli Tan}
\email[Corresponding author: ]{xtan@iastate.edu}
\affiliation{Department of Materials Science and Engineering, Iowa State University, Ames 50011, Iowa, USA}
\author{Lin Zhou}
\email[Corresponding author: ]{linzhou@ameslab.gov}
\affiliation{Ames Laboratory, U.S. Department of Energy, Ames, IA 50011, USA}
\affiliation{Department of Materials Science and Engineering, Iowa State University, Ames 50011, Iowa, USA}

\begin{abstract}
Complex polar structures of incommensurate modulations (ICMs) are revealed in chemically modified PbZrO$_3$ perovskite antiferroelectrics using advanced transmission electron microscopy techniques. The Pb-cation displacements, previously assumed to arrange in a fully-compensated antiparallel fashion, are found to be either antiparallel but with different magnitudes, or in a nearly orthogonal arrangement in adjacent stripes in the ICMs. \textit{Ab initio} calculations corroborate the low-energy state of these arrangements. Our discovery corrects the atomic understanding of ICMs in PbZrO$_3$-based perovskite antiferroelectrics.
 
\end{abstract}

\maketitle

\begin{figure}[h]
\centering
\includegraphics[width=\linewidth]{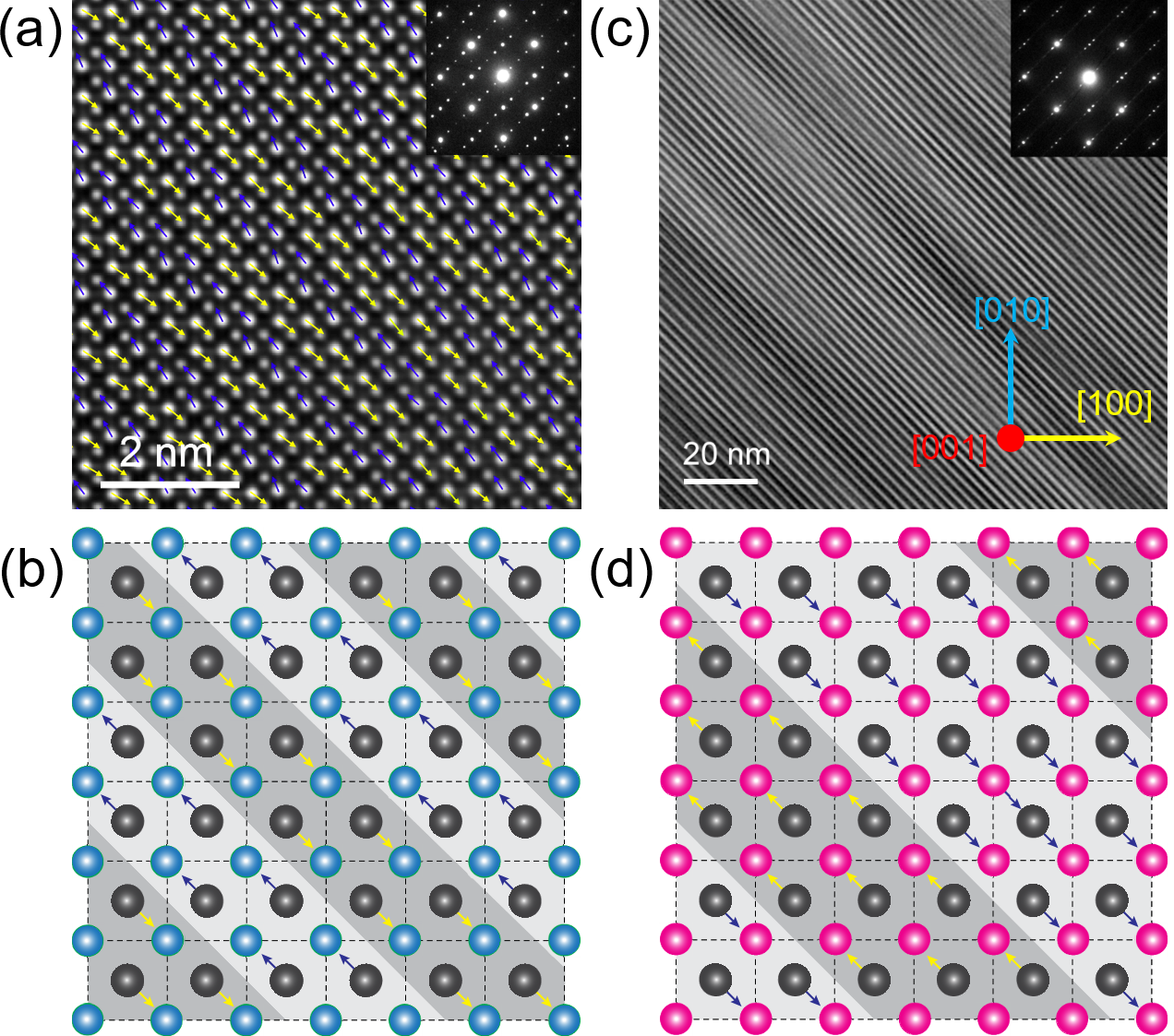} 
\caption{(a) Typical HAADF-STEM image of CMs in PbZrO$_3$, with the SAED pattern shown in the inset. A Pb-displacement vector map is overlaid on the image, showing the antiparallel displacements of Pb-cations in adjacent stripes. (b) Schematic model of the CMs for PbZrO$_3$ proposed by Sawaguchi \textit{et al} \cite{Sawaguchi1951}. A-site (Pb): gray; B-site (Zr): blue. (c) Typical bright-field TEM image of ICMs in a Sn, Ti, Nb co-modified PbZrO$_3$ ceramic, with the SAED pattern shown in the inset. The ICMs render as fine fringes with a periodicity of $\sim$2 nm. (d) Schematic model of the ICMs as a mixture of stripes with a stochastic thickness of 2--4 layers of \{110\}$_c$ plane, proposed by He \textit{et al} \cite{He2005}. A-site (Pb): gray; B-site (Zr, Sn, Ti, Nb): pink.}
\label{fig1}
\end{figure}

Antiferroelectric (AFE) crystals possess dipoles within unit cells in an antiparallel arrangement so that the macroscopic polarization is fully compensated \cite{Kittel1951}. Under the application of an electric field above the critical value, $E_F$, the antiparallel dipoles can be forced to align parallel, leading to the transition to a ferroelectric (FE) phase \cite{Shirane1951}. And presumably, they resume the antiparallel arrangements upon removal of the applied field \cite{Tan2010,Park1997,Tan2011}. Consequently, double hysteresis loops with nearly zero remanent polarizations are recorded from the polarization vs. electric-field (\textit{P-E}) measurement \cite{Park1997,Mani2015}. The most widely studied AFEs are based on PbZrO$_3$, which crystallizes in a distorted perovskite structure with antiparallel Pb-cation displacements. Such a structure is equivalent to a perovskite with commensurate modulations (CMs) of $\frac{1}{4}\{110\}_c$, \textit{i.e.}, a wavelength of four layers of its pseudocubic \{110\}$_c$ plane \cite{Sawaguchi1951,Yamasaki1998,Jona1957,Corker1997}. As shown in Fig. \ref{fig1}(a), the antiparallel arrangement of the Pb-cation displacements in PbZrO$_3$ is directly verified by the mapping of Pb-displacement vectors on a high-angle annular dark-field scanning transmission electron microscopy (HAADF-STEM) image. A schematic diagram \cite{Sawaguchi1951} illustrating the antiparallel arrangement in adjacent stripes is redrawn in Fig. \ref{fig1}(b). 

Because $E_F$ in PbZrO$_3$ at room temperature exceeds its dielectric strength \cite{Fesenko1978}, Sn and Ti in conjunction with a small amount of Nb or La are added to reduce $E_F$ for practical applications \cite{Chang1985,Berlincourt1966,Pan1989,Viehland1995,Viehland1998,Asada2004}. The reduction in $E_F$ is accompanied by the formation of $\frac{1}{n}\{110\}_c$ ($n$ is non-integer, typically between 6 and 8) incommensurate modulations (ICMs), which manifest themselves as fine fringes in diffraction-contrast TEM images and satellite spots in reciprocal space [Fig. \ref{fig1}(c)] \cite{Chang1985,Viehland1995,Viehland1998,Asada2004}. However, the dipole arrangement in these incommensurate phases remains elusive, due largely to their complex sub-domain microstructure \cite{Viehland1995,Viehland1998}. Earlier studies interpreted the ICMs as an ensemble of commensurate phases with coexisting different integer $n$ values (e.g., $n$ = 6, 7, or 8) \cite{Asada2004,He2005,Cai2003}. The dipoles in adjacent stripes were believed to keep a constant magnitude in an antiparallel arrangement across neighboring stripes, so that the overall polarization in AFE domains remain fully-compensated [Fig. \ref{fig1}(d)] \cite{He2005}.

Here we demonstrate that the classic view of the dipole arrangement in ICMs is incorrect. With direct cation displacement mapping in a series of PbZrO$_3$-based compositions (Pb$_{0.99}$Nb$_{0.02}$[(Zr$_{0.57}$Sn$_{0.43}$)$_{1-y}$Ti$_y$]$_{0.98}$O$_3$, referred to as PZ-100$y$ hereafter), we reveal that the Pb-displacements in adjacent stripes of ICMs are not fully compensated most of the time. Instead, they are either antiparallel but with different magnitudes along opposing directions, or in a nearly orthogonal arrangement, \textit{i.e.}, these previously thought fully-compensated AFE domains, in fact, bear a net polarization. These polar AFE domains transition into the FE phase under applied electric fields and resume their original configuration upon removal of the field, leading to the AFE-like double \textit{P-E} hysteresis loops (see Fig. S1 in Supplemental Material \cite{suppl}). 

Figure \ref{fig2} illustrates the representative TEM micrographs of chemically modified PbZrO$_3$ ceramics. Bright-field TEM images in Fig. \ref{fig2}(a), (d) and (g) (top row) from PZ-3, PZ-5, and PZ-6, respectively, show typical 90$^\circ$ AFE domains with textured contrast. Selected area electron diffraction (SAED) patterns along [001]$_c$ zone-axis (shown in the insets) taken from the domain boundary region present the satellite spots, $ha^* + kb^* + lc^* \pm \frac{1}{n}(a^*+b^*)$, ($h$, $k$, $l$ are integer, $n$ represents the periodicity of ICMs indicated in SAED), which are resulted from the ICMs inside each domain \cite{Tan2011}. Atomic-resolution HAADF-STEM images were recorded from the upper and the lower domains of each sample, shown in the second and the third rows, respectively. The atomic positions of both A-site (Pb) and B-site (Zr, Sn, Ti, and Nb) columns were determined by fitting the intensity maxima in these images using a two-dimensional Gaussian function (see Supplemental Material \cite{suppl} for details). The off-center displacements of Pb-cations were calculated with respect to the geometric center of the four surrounding B-site columns and overlaid on the HAADF images.  Strikingly, we found that the dipole arrangement in ICMs deviates from the ``fully-compensated antiparallel model'' depicted in Fig. \ref{fig1}(d). In the upper domains, despite the Pb-displacement vectors are basically antiparallel in adjacent stripes, those in the wider stripes (blue) have larger magnitudes than the ones in the narrower stripes (yellow), as shown in Fig. \ref{fig2}(b), (e), and (h). This imbalance in the magnitudes becomes more obvious from PZ-3 [Fig. \ref{fig2}(b)] to PZ-6 [Fig. \ref{fig2}(h)] when Ti content gets higher. In the lower domains, Fig. \ref{fig2}(c), (f), and (i), the Pb-displacements tend to rotate from the two antiparallel $\langle110\rangle_c$ toward the two orthogonal  $\langle100\rangle_c$ directions. Such rotation becomes more apparent when Ti content increases. In PZ-3, the rotation is mainly noticed in the narrower stripes [yellow in Fig. \ref{fig2}(c)], whereas in PZ-5 almost all the Pb-displacement vectors in the narrower stripes [yellow in Fig. \ref{fig2}(f)] and a large portion of those in the wider stripes [blue in Fig. \ref{fig2}(f)] turned to the $\langle100\rangle_c$ directions. In PZ-6, the Pb-displacements are almost along either [100]$_c$ or $[0\bar{1}0]_c$ [Fig. \ref{fig2}(i)], forming an orthogonal arrangement of dipoles in adjacent stripes. 

\begin{figure*}
\centering
\includegraphics[width=0.9\linewidth]{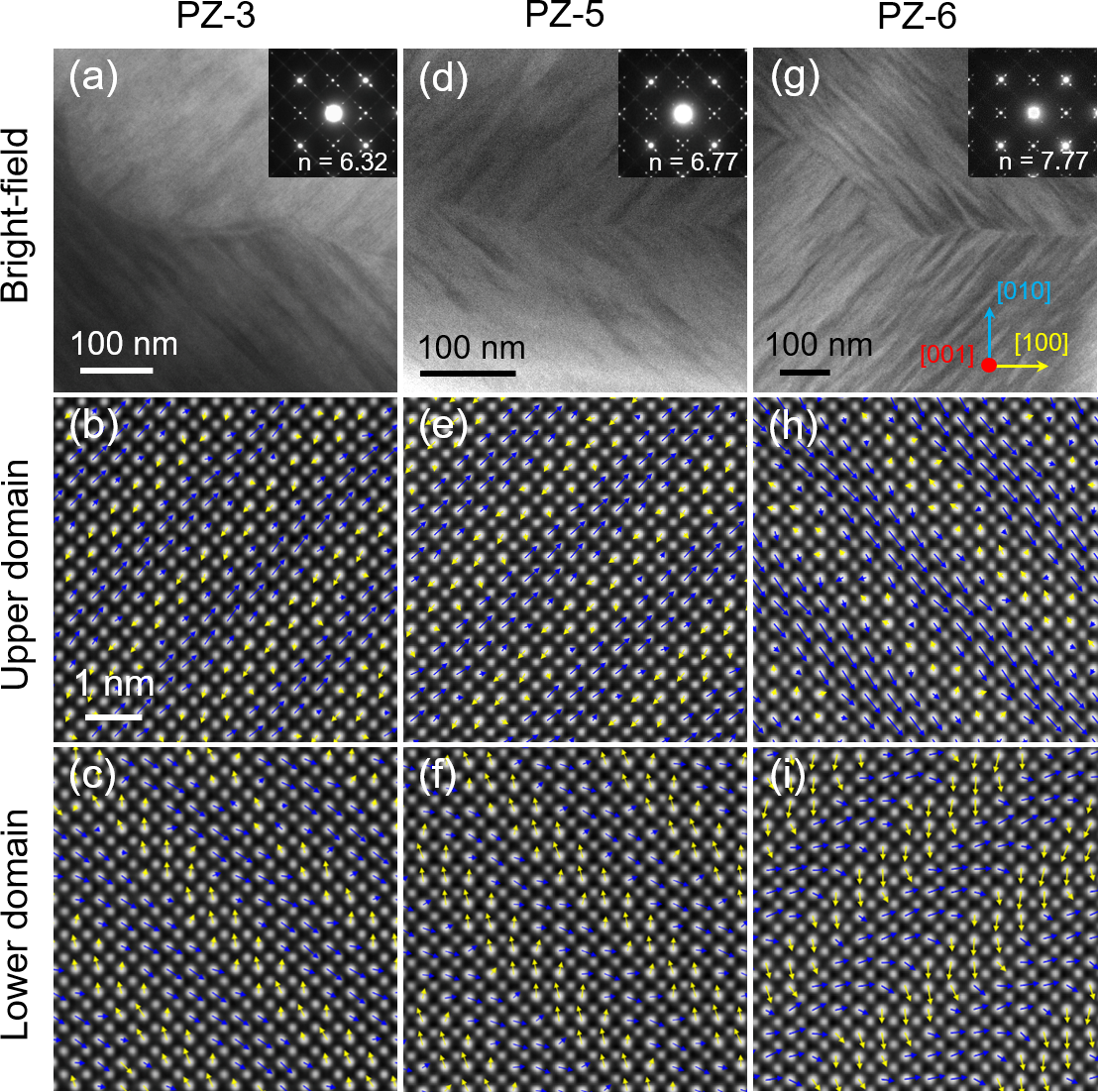} 
\caption{The top row is the typical bright-field TEM images of 90$^\circ$ domains in PZ-3 (a), PZ-5 (d), and PZ-6 (g). The insets are the SAED patterns taken from the domain boundary region showing two sets of satellite spots, from which the periodicity of the ICMs ($n$ value) was calculated. The middle row shows the Pb-displacement vector maps overlaid on the HAADF-STEM images taken from the upper domains in PZ-3 (b), PZ-5 (e), and PZ-6 (h). The Pb-displacements in this row are in an antiparallel arrangement, but those in the wider stripes (blue) have a larger magnitude than the ones in the narrower stripes (yellow). The bottom row presents the Pb-displacement vector maps taken from the lower domains in PZ-3 (c), PZ-5 (f), and PZ-6 (i), where the Pb-displacements tend to form in a nearly orthogonal arrangement in adjacent stripes.}
\label{fig2}
\end{figure*}

We conducted HAADF-STEM image simulations under our experimental conditions. The results confirmed that the observed Pb-displacements reflect the real atomic configuration of dipoles rather than artifacts from the optical system, sample thickness and/or crystal tilts \cite{Liu2017,Cui2017} (see Table S1 and Fig. S2 in Supplemental Material \cite{suppl}). Moreover, HAADF-STEM imaging and atomic-scale energy-dispersive X-ray spectroscopy (EDS) mapping show uniform chemistry across the domain boundaries as well as within the neighboring stripes (see Fig. S3 in Supplemental Material \cite{suppl}), indicating that the observed dipole arrangement is not directly associated with local element distribution. Since the atomic structures found in the composition series are quite distinct from both the classic AFE and FE models, we denote these 90$^\circ$ domains as ``transitional-state'' domains and differentiate the two across the boundary as the ``antiparallel'' side, Fig. 2(b), (e), and (h), and the ``orthogonal'' side, Fig. \ref{fig2}(c), (f), and (i).

\begin{figure*}
\centering
\includegraphics[width=0.8\linewidth]{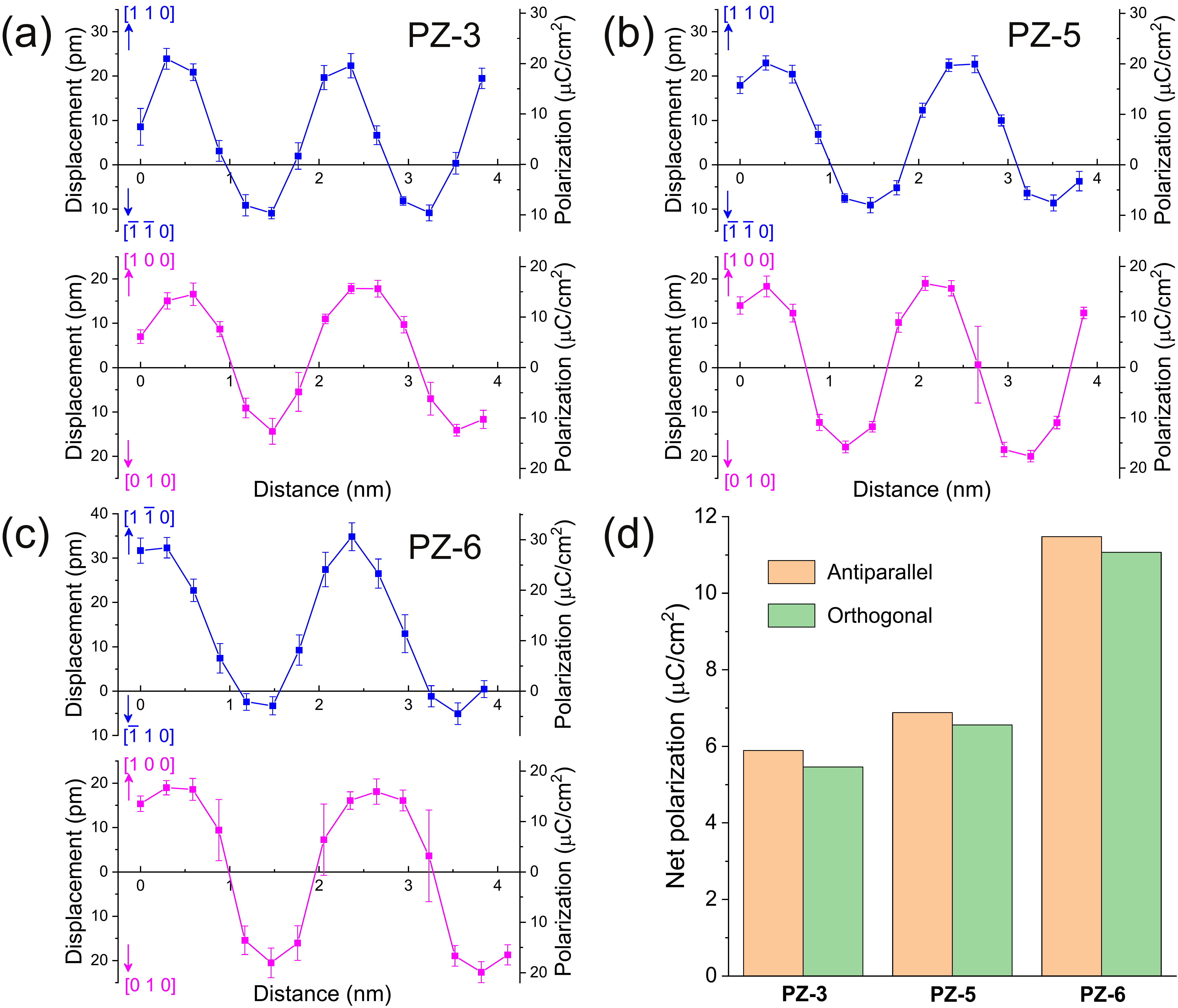} 
\caption{(a--c) Pb-displacement magnitudes, extracted from the Pb-displacement vector maps, and the corresponding polarizations in PZ-3 (a), PZ-5 (b), and PZ-6 (c). The upper plots in each panel (blue) are from the ``antiparallel'' side showing the components along the two antiparallel $\langle110\rangle_c$ directions, whereas the lower plots in each panel (pink) are from the “orthogonal” sides showing the components along the two orthogonal $\langle100\rangle_c$ directions. The displacements were averaged from 10 layers of Pb-cations along the length direction of the ICM stripes. (d) Overall polarizations in the ``antiparallel'' and ``orthogonal'' sides of the three specimens calculated from the corresponding vector maps.}
\label{fig3}
\end{figure*}

To better compare the Pb-displacements in the series with different Ti content, we extract the magnitudes of Pb-displacements from the vector maps in Fig. \ref{fig2} by averaging 10 atomic layers of Pb along the stripes. The corresponding polarization due to Pb-displacements are quantified through the equation:
$$P=\frac{e}{V}\sum_{i=1}^{N}Z^*\cdot\delta_i$$
where $N$ is the number of Pb atoms in the box under consideration, $V$ is the volume of the box, $\delta_i$ is the displacement of $i$th Pb cation, and $Z^*$ is the Born effective charge for Pb cation, given as 3.92 in cubic PbZrO$_3$ by Zhong \textit{et al} \cite{Zhong1994}. These data are plotted against the Pb positions in Fig. \ref{fig3}(a)--(c). It can be seen that the polarization due to Pb-displacements fluctuates across the modulations, rendering wave-like curves. This sinusoidal fashion of the Pb-displacements across the ICMs was previously reported by MacLaren \textit{et al} \cite{MacLaren2012}. However, in the ``antiparallel'' side, these polarization curves (blue) shift toward one side of the two $\langle110\rangle_c$, corresponding to the imbalanced Pb-displacements in opposing directions. In addition, the shift gets more severe as Ti content increases in the composition series. Accompanying with the shift in the ``antiparallel'' side is a slight increase in polarization components along $\langle100\rangle_c$ in the ``orthogonal'' side (pink).

Such unique dipole arrangements in the ICMs lead to uncompensated polarization within these transitional-state domains. The magnitude of the net polarization is estimated by summing the polarization vectors from corresponding vector maps and displayed in Fig. \ref{fig3}(d). The calculated net polarization increases with Ti content in the series, e.g., 5.89 $\mu$C/cm$^2$ and 5.46 $\mu$C/cm$^2$ for the ``antiparallel'' and the ``orthogonal'' side, respectively, in PZ-3; and 11.48 $\mu$C/cm$^2$ and 11.07 $\mu$C/cm$^2$ for the ``antiparallel'' and the ``orthogonal'' side, respectively, in PZ-6. This tendency of increasing net polarization in transitional-state domains, together with the reduced $E_F$ (see Fig. S1 in Supplemental Material \cite{suppl}) in the composition series, confirms that Ti on B-sites in perovskite oxides promotes FE ordering. Furthermore, the presence of uncompensated polarization may be the reason for the fact that all 90$^\circ$ boundaries analyzed in this work have the ``antiparallel'' arrangement on one side and the ``orthogonal'' arrangement on the other side. A 90$^\circ$  boundary with the same dipole arrangement on both sides would carry a very high depolarization energy.

\begin{figure}
\centering
\includegraphics[width=\linewidth]{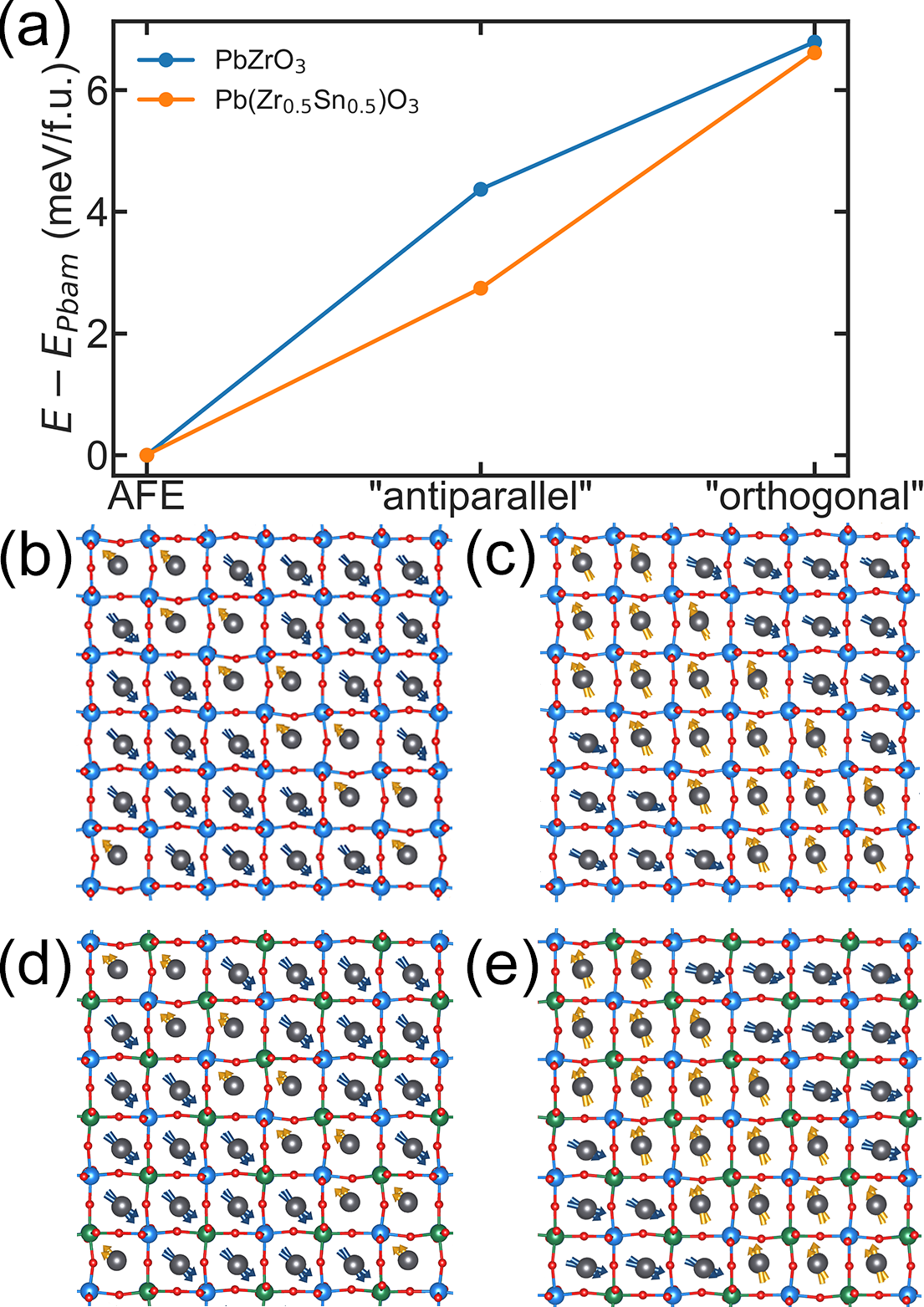} 
\caption{(a) Energies of the ``antiparallel'' and ``orthogonal'' domain structures relative to the AFE phase ($Pbam$-like) for PbZrO$_3$ and Pb(Zr$_{0.5}$Sn$_{0.5}$)O$_3$. (b, c) DFT optimized ``antiparallel'' and ``orthogonal'' domain structures of PbZrO$_3$. (d, e) DFT optimized ``antiparallel'' and ``orthogonal'' domain structures of Pb(Zr$_{0.5}$Sn$_{0.5}$)O$_3$. Pb: gray; Zr: blue; Sn: green; O: red. Pb-displacements are denoted in blue and yellow colors in different stripes. The atom position and displacement data used to create (b), (c), (d), and (e) are listed in Table S4, S5, S6, and S7, respectively, in the Supplemental Material \cite{suppl}.}
\label{fig4}
\end{figure}

More insight on the formation of the ``antiparallel'' and ``orthogonal'' domains was gained from density functional theory (DFT) calculations (at 0 K). To assess the stability of these polar domain structures, they were studied for both PbZrO$_3$ and Pb(Zr$_{0.5}$Sn$_{0.5}$)O$_3$ and compared with the AFE phase ($Pbam$-like), the ground state of PbZrO$_3$. Note that Pb(Zr$_{0.5}$Sn$_{0.5}$)O$_3$ is considered to largely resemble the experimental compositions at a feasible computational cost. The DFT optimized “antiparallel” and “orthogonal” domain structures for PbZrO$_3$ and Pb(Zr$_{0.5}$Sn$_{0.5}$)O$_3$ are displayed in Fig. \ref{fig4}(b)--(e), and their ground state structures ($Pbam$-like) are present in Fig. S4(a) and S4(b). Note that the Pb-displacements obtained by DFT are mostly in the (001)$_c$ plane with zero or negligible out-of-plane components (see detailed atom position and displacement data in Table S2--S7 in the Supplemental Material \cite{suppl}). For the relaxed transitional-state domain structures in pure PbZrO$_3$ as shown in Fig. \ref{fig4}(b) and \ref{fig4}(c), they retain the characteristics of the ``antiparallel'' and ``orthogonal'' domains found in the HAADF-STEM images, and their energies are only slightly higher than the AFE phase [Fig. \ref{fig4}(a)]. The ``antiparallel'' structure has an energy of 4.3 meV/f.u. higher than AFE, and it is 6.8 meV/f.u. for the ``orthogonal'' structure. Similarly, the ``antiparallel'' and ``orthogonal'' structures are also stable in  Pb(Zr$_{0.5}$Sn$_{0.5}$)O$_3$, as shown in Fig. \ref{fig4}(d) and \ref{fig4}(e); and compared with PbZrO$_3$ the relative energies are only 2.8 and 6.7 meV/f.u. higher than AFE. A close examination of the ``orthogonal'' structure reveals that the Pb-displacements in the ``blue'' and ``yellow'' stripes in  Pb(Zr$_{0.5}$Sn$_{0.5}$)O$_3$  are aligned closer to the [100]$_c$ and [010]$_c$ directions, respectively, than in PbZrO$_3$. The optimized structures and the small energy differences indicate that the ``antiparallel'' and ``orthogonal'' domain structures are meta-stable, and the stability improves with the substitution of Sn on the B-sites. Note that the transitional-state domain structures are more complex than the AFE phase; therefore, they are likely to be more favorable at finite temperature due to the entropy contribution to the free energy. Furthermore, the stabilization of the transitional-state domain structures is related to the existence of a bi-linear coupling between the Pb-displacements and the oxygen octahedral tilting \cite{Patel2016,Xu2019}, such that the octahedra develop an accommodating pattern to lower the energy cost of forming Pb-modulations (details are discussed in Supplemental Material \cite{suppl}).

The presence of net polarization in the transitional-state domains appears to contradict the measurement of double \textit{P-E} hysteresis loops on bulk samples in this composition series (Fig. S1 in Supplemental Material \cite{suppl}). To reconcile this conflict, we performed \textit{in-situ} electric-biasing TEM experiments. The results show the transition of a transitional-state domain to the FE phase under an applied electric field, and its resumption to the original transitional-state domain structure upon the removal of the field (see Fig. S5 in Supplemental Material \cite{suppl} for details). The full recovery of the hierarchical structure (domains, boundary, ICMs, and cation-displacements) hence explains the nearly zero remnant polarization measured on bulk samples. 

In summary, we uncover at the atomic-scale that the Pb-displacements in ICMs of PbZrO$_3$-based AFEs are either antiparallel but imbalanced or in a nearly orthogonal arrangement. The uncompensated dipole moments introduce non-zero net polarizations in 90$^\circ$ AFE domains, and their magnitudes increase with Ti content. In contrast to the polar nature of the microstructure, macroscopically these ceramics behave in an AFE manner with the characteristic \textit{P-E} double hysteresis loops of negligible remnant polarization. Such unique structures in incommensurate modulated PbZrO$_3$-based AFEs can be viewed as a transitional state between long-range FE and AFE orderings. Experiments on other inorganic AFE oxides may also find such transitional state in the future.

This work was supported by the National Science Foundation (NSF) through Grant DMR-1700014. All electron microscopy work was performed at the Sensitive Instrument Facility at Ames Laboratory, which is operated for the U.S. DOE by Iowa State University under Contract No. DE-AC02-07CH11358. B.X. acknowledges funding from Air Force Office of Scientific Research (AFOSR) under Grant FA9550-16-1-0065. L.B. acknowledges support from Office of Naval Research (ONR) under Grant N00014-17-1-2818.

\end{document}


\title{Supplemental Material for\\Uncompensated Polarization in Incommensurate Modulations of Perovskite Antiferroelectrics}%
\author{Tao Ma}%
\thanks{These authors contributed equally to this work.}
\affiliation{Ames Laboratory, U.S. Department of Energy, Ames, IA 50011, USA}%
\author{Zhongming Fan}
\thanks{These authors contributed equally to this work.}
\affiliation{Department of Materials Science and Engineering, Iowa State University, Ames 50011, Iowa, USA}
\author{Bin Xu}
\affiliation{Physics Department and Institute for Nanoscience and Engineering, University of Arkansas, Fayetteville, Arkansas 72701, USA}
\affiliation{School of Physical Science and Technology, Soochow University, Suzhou, Jiangsu 215006, China}
\author{Tae-Hoon Kim}
\affiliation{Ames Laboratory, U.S. Department of Energy, Ames, IA 50011, USA}%
\author{Ping Lu}
\affiliation{Sandia National Laboratories, PO Box 5800, MS 1411, Albuquerque, NM 87185, USA}
\author{Laurent Bellaiche}
\affiliation{Physics Department and Institute for Nanoscience and Engineering, University of Arkansas, Fayetteville, Arkansas 72701, USA}
\author{Matthew J. Kramer}
\affiliation{Ames Laboratory, U.S. Department of Energy, Ames, IA 50011, USA}
\affiliation{Department of Materials Science and Engineering, Iowa State University, Ames 50011, Iowa, USA}
\author{Xiaoli Tan}
\email[Corresponding author: ]{xtan@iastate.edu}
\affiliation{Department of Materials Science and Engineering, Iowa State University, Ames 50011, Iowa, USA}
\author{Lin Zhou}
\email[Corresponding author: ]{linzhou@ameslab.gov}
\affiliation{Ames Laboratory, U.S. Department of Energy, Ames, IA 50011, USA}
\affiliation{Department of Materials Science and Engineering, Iowa State University, Ames 50011, Iowa, USA}

{\large\noindent Supplemental Material for}\\[1em]
\begin{center}
\large\bf Uncompensated Polarization in Incommensurate Modulations of Perovskite Antiferroelectrics
\end{center}

\section{Experimental details}
\subsection{Ceramic preparation} 
The solid-state reaction method was used to prepare PbZrO$_3$ and Pb$_{0.99}$Nb$_{0.02}$[(Zr$_{0.57}$Sn$_{0.43}$)$_{1-y}$Ti$_y$]$_{0.98}$O$_3$ ($y = 0.03, 0.05, 0.06$) polycrystalline ceramics, using PbO ($\geq$ 99.9\%), Nb$_2$O$_5$ ($\geq$ 99.999\%), ZrO$_2$ ($\geq$ 99.99\%), SnO$_2$ ($\geq$ 99.997\%), and TiO$_2$ ($\geq$ 99.9\%) as the starting materials. The raw powders were mixed in the stoichiometric amounts (5\% excess PbO) and vibratorily milled in ethanol with zirconia mill media for 6 h. The mixture was then dried and calcined at 850 $^\circ$C for 4 h. With polyvinyl alcohol as binder, the calcined powder was uniaxially pressed into circular disks under 300 MPa. Buried in a protective powder of the same composition, the disks were sintered at 1250 $^\circ$C for 2 h for PbZrO$_3$ and 1300 $^\circ$C for 3 h for modified compositions, respectively, in alumina crucibles at a ramp rate of 5 $^\circ$C/min.

\subsection{Electrical property measurement} Silver electrodes were painted and fired on the polished surfaces of sintered pellets at 800 $^\circ$C for 6 min. The polarization vs. electric field (\textit{P-E}) hysteresis loops were measured using a standardized ferroelectric test system (Precision LC II, Radiant Technologies) at 1 Hz. 

\subsection{Transmission electron microscopy} TEM observation was performed using an FEI Titan Themis 300 probe-corrected STEM with an accelerating voltage of 200 kV. TEM specimens were prepared by mechanical wedge-polishing, followed by a short time, low-voltage Ar ion-mill. A thin layer of carbon was deposited on the specimens prior to the observation to avoid charging. High-resolution HAADF images were captured using a sub-angstrom electron probe with a convergence angle of 18 mrad and a detection angle of 99--200 mrad. The electron beam current was kept around 20 pA. The orientation of the grain was carefully aligned to the [001]$_c$ zone-axis by monitoring the Ronchigram. To minimize the beam-specimen interaction, scanning distortion, and readout noise, 40 images were recorded from each area with a frame rate of 0.2 s, and then cross-correlated and summed to produce a drift-corrected frame-integrated image using a built-in function of Velox software (Thermo Fisher Scientific). 

\subsection{Calculation of the Pb-displacement vector map} The image post-processing workflow to generate the Pb-displacement vector map is written in \textit{Python} 3.6. First, the atomic-resolution image is loaded into the workspace using \textit{Hyperspy} \cite{Pena2018} package, and converted to a 2D array of intensity data. Then, \textit{Atomap} \cite{Nord2017} package is used to find the initial positions of Pb atomic columns. These initial positions are passed into the center-of-mass refining function to get approximate positions of atoms, which are used as the initial input for the second round refining using 2D Gaussian function. After that, we use a 2D Gaussian-based function in \textit{Atomap} to remove the Pb atomic columns from the original image. The modified image, which contains only B-site atomic columns, is used to find B-site atom positions via the same procedure. The Pb-displacement vector is calculated by comparing the actual Pb positions to the mathematical center of the quadrilateral formed by the four surrounding B-site cations. By iteratively performing the same procedures on each Pb column, a vector map is generated. The displacement vectors are drawn on each Pb column in the original image, with the arrowheads pointing to the displacement directions, and lengths representing the magnitude by a scaling factor. 

\subsection{HAADF-STEM image simulation} A series of image simulation on PbZrO$_3$ at different specimen thickness and crystal tilts was carried out using QSTEM software based on a frozen phonon multislice technique \cite{Koch2002}. The microscope aberration coefficients used for simulation (listed in Table S1) are the largest acceptable values measured by the DCOR+ software before we proceed with the experiments. The simulation was performed in a 23.5 \AA $\times$ 23.5 \AA\ (containing 5 $\times$ 5 cubic unit cell) region with 100 $\times$ 100 sampling pixels, giving a resolution of 23.5 pm/pixel that is comparable with the actual images (22.6 pm/pixel under the magnification of 3.6 MX). PbZrO$_3$ [001]$_c$ slabs with a thickness of 20 nm, 40 nm, 60 nm, and 80 nm were constructed for image simulation. The upper bound thickness, 80 nm, was determined by the electron energy loss spectroscopy (EELS) taken at an area thicker than the imaging regions based on a log-ratio method (the inelastic mean free path was estimated to be 80 nm). Crystal tilts were applied along both $\alpha$ and $\beta$ directions in the simulation, in a geometry shown in Fig. S2c. 

\subsection{Atomic-scale energy-dispersive X-ray spectroscopy (EDS) mapping} The EDS spectra were acquired using a SuperX EDX detector, with a probe convergence angle of 18 mrad. A beam current of $\sim$30 pA was used during acquisition to reduce radiation damage to the specimen while maintaining reasonable X-ray counts. The spectral images, 400 $\times$ 400 pixels (6.25 $\times$ 6.25 nm), were recorded as a series of frames with real-time software (Bruker Esprit) drift correction and precise sample movement using a piezo-stage. The instantaneous dwell time on each pixel was 8 $\mu$s, giving a frame time of 1.28 s. The total acquisition time was $\sim$20 min, yielding an equivalent per-pixel dwell time of about 8 ms. The experimental conditions were optimized to reduce possible radiation damage during the acquisition. Atomic-scale HAADF images were taken before and after the EDS mapping to verify there was no visible radiation damage induced during the acquisition (images not shown). To increase the signal-to-noise ratio of the EDS maps (this reduces the total acquisition time necessary), the spectral images were integrated along the stripe direction ([110]$_c$) using a lattice-averaging method \cite{Lu2014}. The lattice average did not alter the composition across the stripes since it was only done in the direction parallel to the stripe plane. 

\subsection{\textit{In-situ} electric-biasing TEM} Disk specimens with 3 mm in diameter were prepared from as-sintered pellets through standard procedures, including grinding, cutting, dimpling, and ion mill to perforation. After connecting the sample to a dedicated specimen holder, \textit{in-situ} electric-biasing TEM experiments were carried out using a Tecnai G2-F20 microscope operated at 200 kV.
  
\section{Computational details}
The $Pbam$ structure of PbZrO$_3$ and Pb(Zr$_{0.5}$Sn$_{0.5}$)O$_3$ was simulated with a $\sqrt{2} \times 2\sqrt{2} \times 2$ (40 atoms) cell; domain 1 and domain 2 structures, constructed analogous to the ``antiparallel'' and ``orthogonal'' domains, were simulated with $\sqrt{2} \times 3\sqrt{2} \times 2$ (60 atoms) and $\sqrt{2} \times 4\sqrt{2} \times 2$ (80 atoms) supercells, respectively. Note that an approximation composition of Pb(Zr$_{0.5}$Sn$_{0.5}$)O$_3$ was considered for the chemically modified case. Sn composition of 0.5 is used here, which is close to the experimental value, and it allows us to adopt supercell sizes that are affordable for \textit{ab initio} calculations. Chemical modification from small concentrations of Nb and Ti is neglected, which should not affect the interpretation of the results. For Pb(Zr$_{0.5}$Sn$_{0.5}$)O$_3$, a rocksalt arrangement of Zr and Sn atoms was incorporated, which can approximate the uniform distribution of B-site cations and capture the local structure characteristics. DFT calculations were performed using the Vienna \textit{ab-initio} simulation package (VASP) \cite{Kresse1999}. The projected augmented wave (PAW) potentials are used, with the generalized gradient approximation (GGA) and the Perdew-Burke-Ernzerhof (PBE) exchange-correlation functional for solid (PBESol) \cite{Perdew2008}. The plane wave cutoff energy is 500 eV. Pb 6s6p, Zr 4s4p4d5s, Sn 4d5s, and O 2s2p states were treated as valence electrons. $6 \times 4 \times 4$ and $6 \times 2 \times 4$ Monkhorst-Pack $k$-meshes were used for the $Pbam$ and domain structures, respectively, and the structures are fully optimized until all ionic forces in the system were within 0.001 eV/\AA. Detailed Pb position and displacement data in the DFT optimized domain structures are attached in Table S2--S7. These data were used to create Fig. 4(b)--(d) and Fig S4.

\section{Correlation between lead-displacements and the antiferrodistortive distortions}
It is well known that in many perovskites the electric dipoles are coupled to the oxygen octahedral tiltings, which is also termed antiferrodistortive (AFD) distortions. In particular, a special bi-linear coupling between local displacement and AFD tiltings has been recently proposed to play an important role to stabilize the ground state $Pbam$ phase of PbZrO$_3$ \cite{Patel2016}. We numerically find that the DFT relaxed structures also exhibit an apparent correlation between the Pb-displacements and the AFD pattern, which may be related to the formation of the transitional-state domain structures. For the AFE $Pbam$-like structure, the Fourier transform (FT) of the in-plane components of the Pb displacements yields the $\Sigma$ point $\frac{2\pi}{a}(\frac{1}{4},\frac{1}{4},0)$, and FT of the AFD tiltings yields the same $\Sigma$ point  $\frac{2\pi}{a}(\frac{1}{4},\frac{1}{4},0)$, for the $z$ (out-of-plane) component (as consistent with the aforementioned bilinear coupling \cite{Patel2016}) and the R point  $\frac{2\pi}{a}(\frac{1}{2},\frac{1}{2},\frac{1}{2})$ for the in-plane components. Similar correlations are also seen in the relaxed ``antiparallel'' and ``orthogonal'' structures, for instance, FT of  $\frac{2\pi}{a}(\frac{1}{6},\frac{1}{6},0)$, and  $\frac{2\pi}{a}(\frac{1}{8},\frac{1}{8},0)$ (that are thus along the $\Gamma$--$\Sigma$ line) is predominant for the in-plane Pb-displacements and the $z$-component of AFD tiltings for the ``antiparallel'' and ``orthogonal'' case, respectively, (which is once again fully in-line with the bilinear coupling discovered in Ref. \cite{Patel2016}) while the R point is most prominent for the $z$-component of Pb-displacements and in-plane components of AFD tiltings in both cases. 

\section{Supplemental figures}

\begin{figure}[h]
\centering
\includegraphics[width=0.6\linewidth]{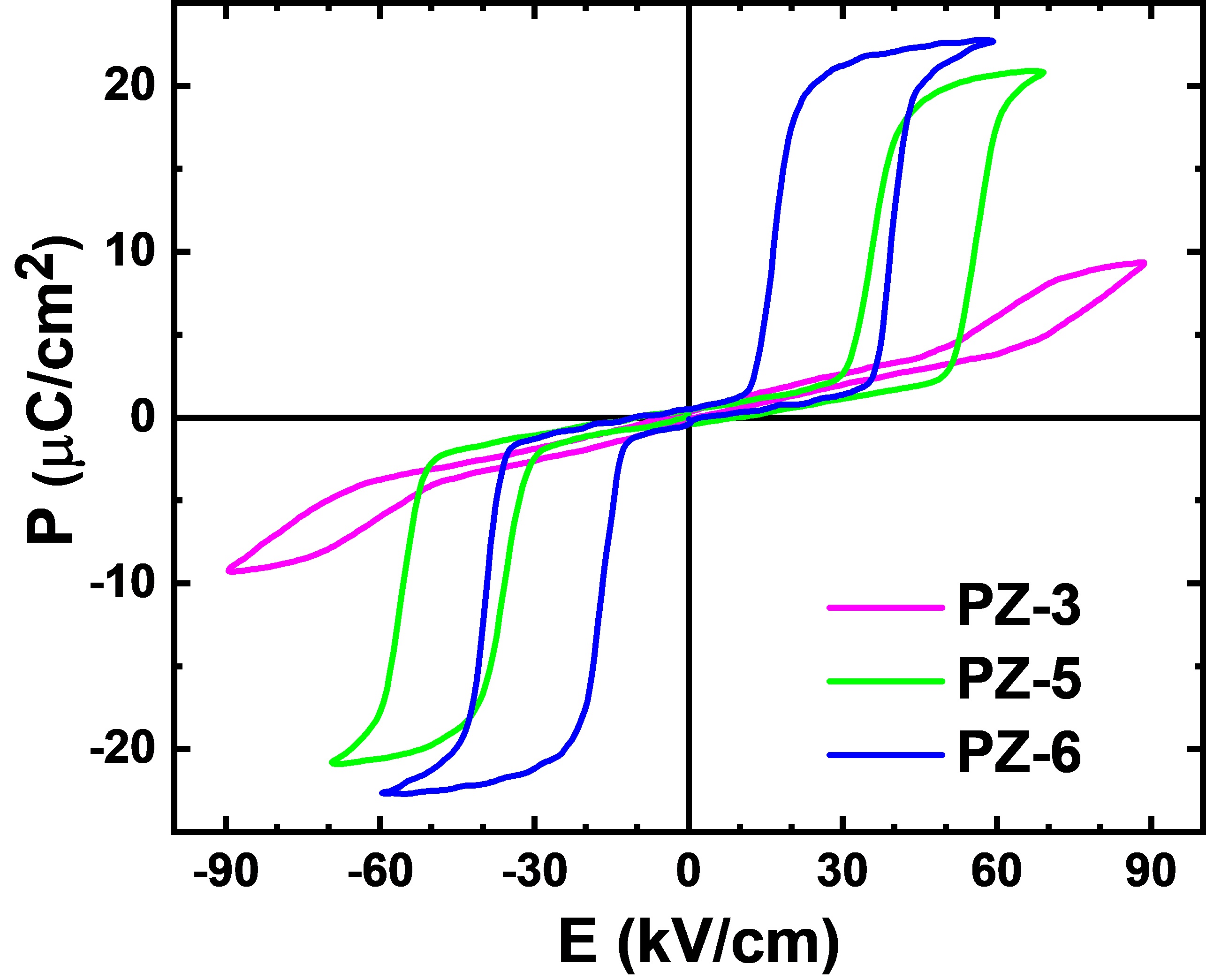} 
\caption{The \textit{P-E} hysteresis loops measured from bulk specimens of PZ-100$y$ ($y = 0.03, 0.05, 0.06$) at 4 Hz at room temperature. All specimens display the characteristic double hysteresis loops with nearly zero remanent polarization. As Ti content increases, the critical field ($E_F$) for phase transition decreases.}
\end{figure}

\begin{figure}[h]
\centering
\includegraphics[width=0.8\linewidth]{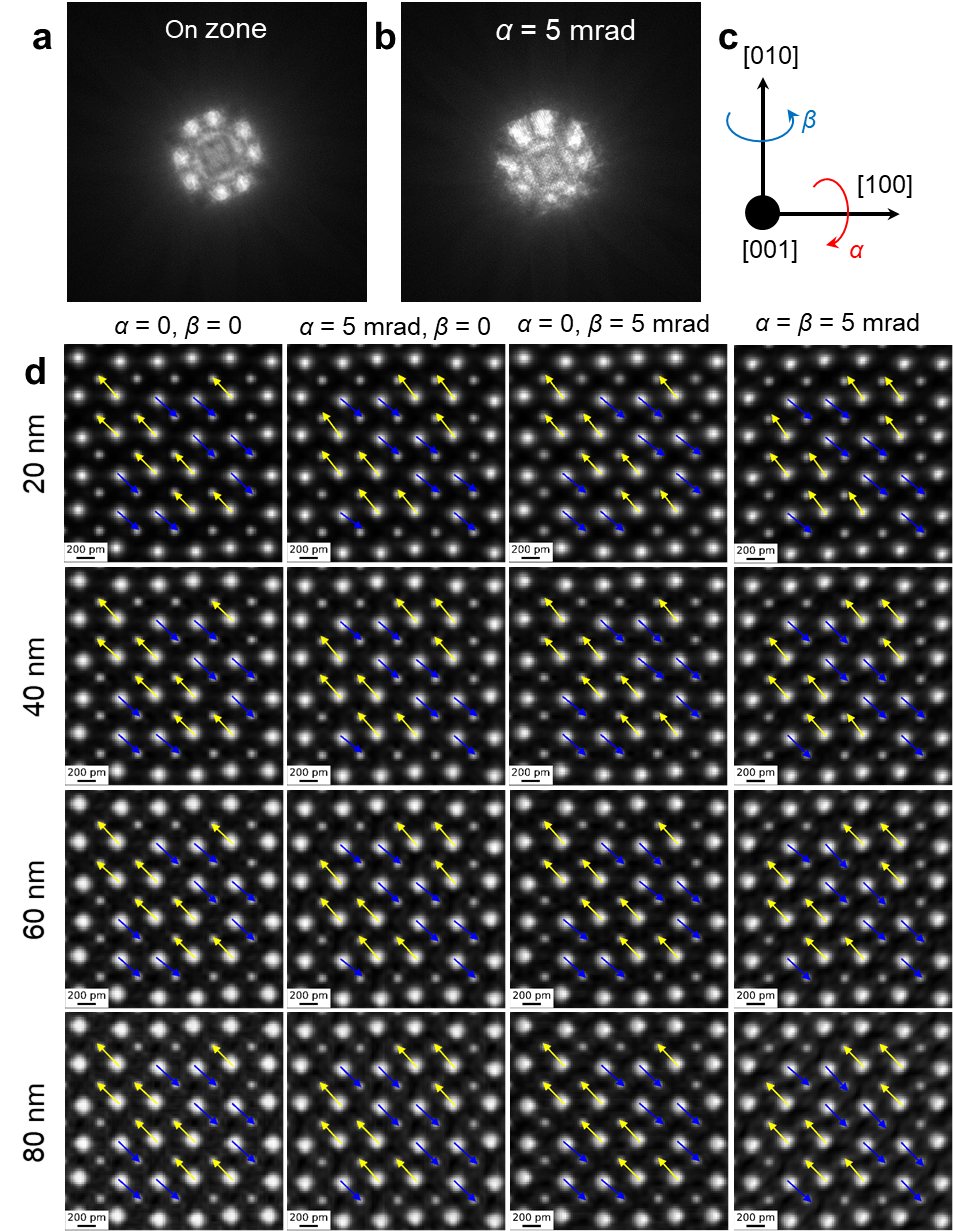} 
\caption{Specimen thickness and tilt effects on the determination of Pb displacements.  (a, b) Electron Ronchigrams of the specimen oriented along [001]$_c$ zone axis (a) and tilted at $\alpha = 5$ mrad off from [001]$_c$ (b). An off-tilt of only 5 mrad already results in an obviously visible asymmetry in the Ronchigram and was therefore set as the upper bound for the crystal tilts. (c) Tilting geometry used for image simulation. (d) Simulated HAADF-STEM images of PbZrO$_3$ [001]$_c$ at different thickness and tilts, overlaid with Pb-displacement vector maps. The antiparallel arrangement of Pb displacement, as expected from the input model, is revealed in all the simulated images. Even in the extreme case, an 80 nm thick PbZrO$_3$ tilted 5 mrad in both $\alpha$ and $\beta$ directions, we still resolved the antiparallel Pb displacement expected for PbZrO$_3$ without observable artifacts. The results confirmed that our determination of dipole arrangements from the atomic resolution HAADF-STEM images is technically reliable, and can reflect the real atomic configurations of the sub-domain superstructures in PbZrO$_3$-based antiferroelectrics.}
\end{figure}
 
\begin{figure}[h]
\centering
\includegraphics[width=0.6\linewidth]{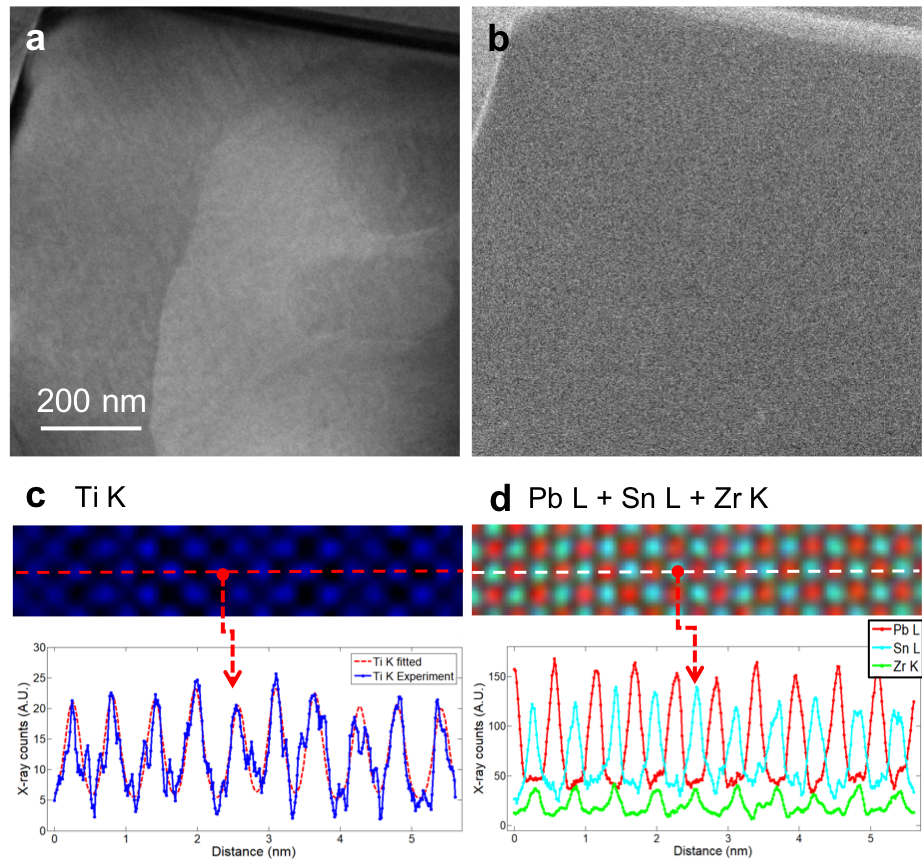} 
\caption{Evaluation of chemistry variation in the 90$^\circ$ domains of PZ-5. (a) Bright-field TEM image of 90$^\circ$ domains showing typical textured contrast due to the appearance of ICMs. (b) HAADF-STEM image of the same region. The uniform contrast indicates no detectable chemistry variation in the domain scale due to the $Z$-contrast. (c, d) Lattice-averaged EDS maps across the ICMs for Ti K (c) and Pb L (red) + Sn L (cyan) + Zr K (green) (d), along with the intensity line-profiles extracted from the dashed lines. The lattice-average was done only in the direction parallel to the ICMs so it did not alter the composition across the ICMs. The uniform intensity on each atomic column confirmed the homogeneous distribution of Pb (on the A-site), Sn, and Zr (on the B-site) over the ICMs, the sub-domain scale. }
\end{figure}

\begin{figure}[h]
\centering
\includegraphics[width=0.8\linewidth]{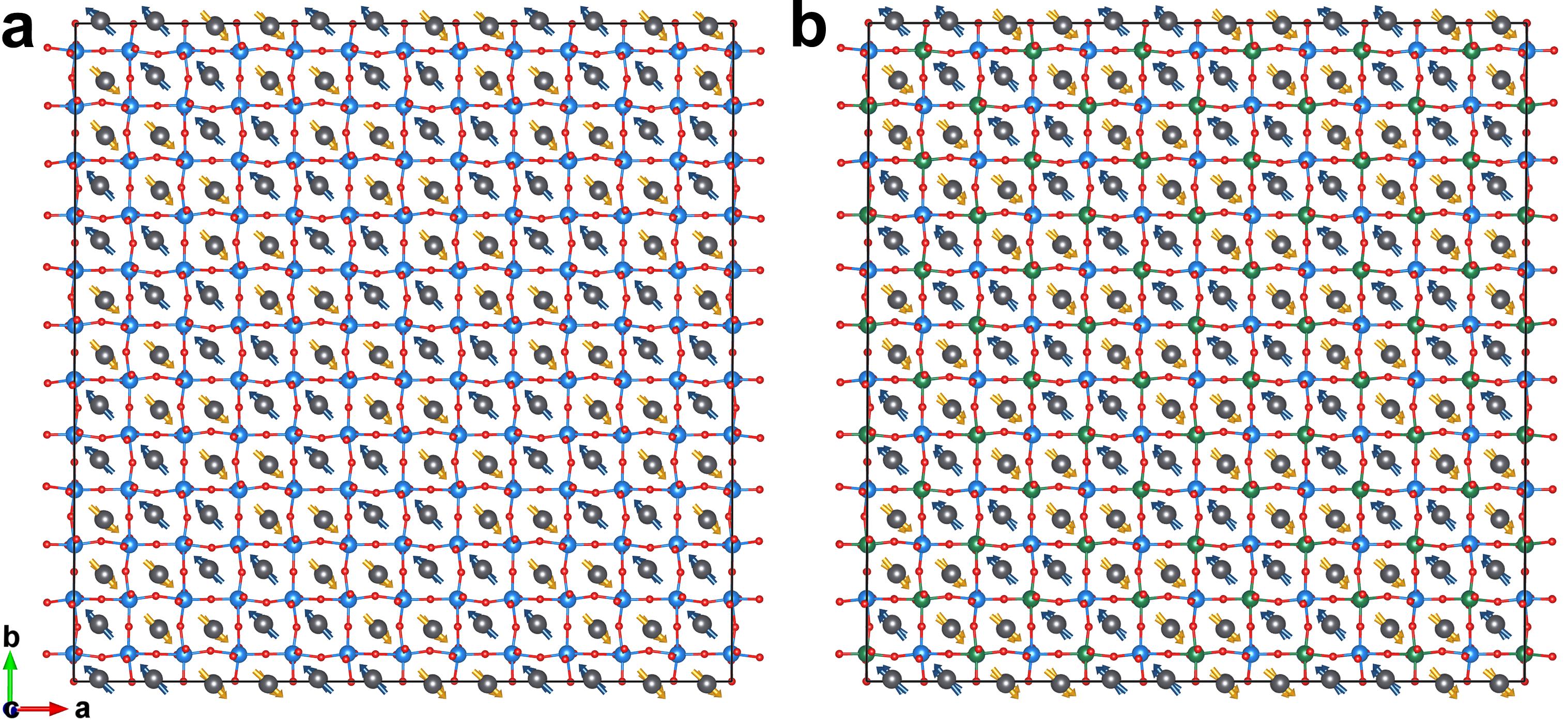} 
\caption{DFT optimized domain structures of the AFE phase ($Pbam$-like) for (a) PbZrO$_3$ and (b) Pb(Zr$_{0.5}$Sn$_{0.5}$)O$_3$. Pb: gray; Zr: blue; Sn: green; O: red. Pb-displacements are denoted in blue and yellow colors in different stripes. The atom position and displacement data used in (a) and (b) were listed in Table S2 and S3, respectively.}
\end{figure} 

\begin{figure}[h]
\centering
\includegraphics[width=0.8\linewidth]{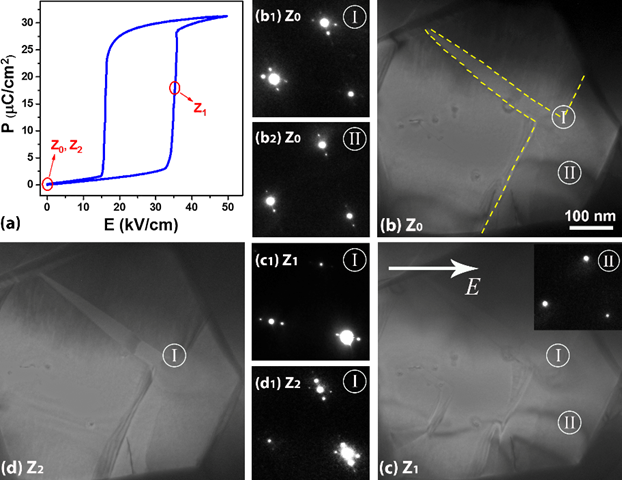} 
\caption{Direct observation of the response of the transitional-state domains to an applied electric field in PZ-6. (a) \textit{P-E} loop under one unipolar cycle. The red open circles on the loop marked as Z$_0$–Z$_2$, correspond to the levels of applied fields at which bright-field images and SAED were recorded. (b–d) Bright-field micrographs at the applied fields corresponding to Z$_0$–Z$_2$, respectively. The grain of interest was viewed near [111]$_c$, with two transitional-state domains present in the virgin state. The yellow dashed line in (b) highlights the 90$^\circ$ domain boundary. Two sets of satellite spots are present in the SAED taken from I (b$_1$), whereas only one set is present from II (b$_2$). When an electric field was applied with a direction indicated by the bright arrow in (c) and a level corresponding to Z$_1$ in (a), the lower domain transformed to the FE phase evidenced by the disappearance of the satellite spots in the SAED \cite{Guo2015} (inset of c), whereas the upper domain persisted, still giving the satellite spots in the SAED (c$_1$). After the field was removed, the lower domain resumed to its original state, revealed by the reappearance of the satellite spots in the SAED (d$_1$).}
\end{figure} 
 
\clearpage
\section{Supplemental Table}
\renewcommand{\arraystretch}{2}
\begin{table}[h]
\centering
\caption{Parameters used for HAADF-STEM image simulation.}
\begin{tabularx}{\textwidth}{lllX}
\hline
Voltage \hspace{2em}  & Detection range  \hspace{2em} & Convergence angle \hspace{2em}   & Aberrations\\
\hline
200 kV & 99--200 mrad & 18 mrad & A1 = 2 nm; C1 = 0 nm\newline A2 = 75 nm; B2 = 50 nm \newline A3 = 1 $\mu$m; S3 = 1 $\mu$m; C3 = 1 $\mu$m \newline A4 = 10 $\mu$m; D4 = 10 $\mu$m; B4 = 10 $\mu$m \newline A5 = 1 mm; R5 = 1 mm; S5 = 1 mm; C5 = 1 mm\\
\hline
\end{tabularx}
\end{table}

\begin{table}[h]
\centering
\caption{DFT optimized Pb positions ($x, y, z$) and displacements ($D_x, D_y, D_z$) in the ground state ($Pbam$) of PbZrO$_3$. The displacement data were calculated by the difference between the relaxed structure and the highly symmetric (cubic) structure.}
\begin{tabularx}{\textwidth}{lXXXXXX}
\hline
Site \hspace{2em} & $x$ (\AA) & $y$ (\AA) & $z$ (\AA) & $D_x$ (\AA) & $D_y$ (\AA) & $D_z$ (\AA)\\
\hline
Pb1 &     1.855029685991 &     0.204107916343 &     0.000000000000 &    -0.221041536572 &     0.201574440015  &    0.000000000000\\
Pb2&     10.606464697164&      3.963235075457 &     0.000000000000  &    0.221041826655 &    -0.201574904700  &    0.000000000000\\
Pb3 &    6.434855284810 &     3.938701528811 &     0.000000000000  &    0.201574631206 &    -0.221041378558  &    0.000000000000\\
Pb4&   6.026639443809  &    0.228642148514   &   0.000000000000 &    -0.201573995660  &    0.221041599396 &     0.000000000000\\
Pb5&      1.932947568603   &   0.186082011365  &    4.080414262505  &   -0.143123653960  &    0.183548535037 &     0.000000000000\\
Pb6&     10.528546483009   &   3.981261494896   &   4.080414262505   &   0.143123612500  &   -0.183548485261   &   0.000000000000\\
Pb7&     6.416828955796&      4.016618987387  &    4.080414262505  &    0.183548302192  &   -0.143123919981 &     0.000000000000\\
Pb8&      6.044665514873 &     0.150724266105   &   4.080414262505 &    -0.183547924596  &    0.143123716987  &    0.000000000000\\
\hline
\end{tabularx}
\end{table}

\begin{table}[h]
\centering
\caption{DFT optimized Pb positions ($x, y, z$) and displacements ($D_x, D_y, D_z$) in the ground state ($Pbam$-like) of Pb(Zr$_{0.5}$Sn$_{0.5}$)O$_3$. The displacement data were calculated by the difference between the relaxed structure and the highly symmetric (cubic) structure.}
\begin{tabularx}{\textwidth}{lXXXXXX}
\hline
Site \hspace{2em} & $x$ (\AA) & $y$ (\AA) & $z$ (\AA) & $D_x$ (\AA) & $D_y$ (\AA) & $D_z$ (\AA)\\
\hline
Pb1&   1.872461335363  &    0.157021228712 &     0.002936145811 &    -0.184700595257  &    0.153714150459  &    0.003311916844\\
Pb2&   10.477124499000 &     3.977144793627 &    -0.005942314034  &    0.184700882519 &    -0.153714610800  &   -0.003311916813\\
Pb3 &      6.331814321305 &     3.939544690360  &    0.001433061605  &    0.153714339969  &   -0.184700438609  &    0.003311916781\\
Pb4&      6.017771855639  &    0.194622011392 &    -0.004439230014   &  -0.153713710127  &    0.184700657680   &  -0.003311916936\\
Pb5&      1.942647924643 &     0.176257984204  &    4.063164414953  &   -0.113769269901  &    0.173695642027 &    -0.006046140081\\
Pb6&     10.405448109242  &    3.956419075630 &     4.073002068926  &    0.113769228837  &   -0.173695592721   &   0.006046140080\\
Pb7&      6.351050656461 &     4.009730859304   &   4.061661330901  &    0.173695411201 &    -0.113769533588 &    -0.006046139991\\
Pb8&      5.997045792568  &    0.122945950224 &     4.074505152948     &-0.173695037122  &    0.113769332588 &     0.006046139960\\
\hline
\end{tabularx}
\end{table}

\begin{table}[h]
\centering
\caption{DFT optimized Pb positions ($x, y, z$) and displacements ($D_x, D_y, D_z$) in the ``antiparallel'' domain structure of PbZrO$_3$. The displacement data were calculated by the difference between the relaxed structure and the highly symmetric (cubic) structure.}
\begin{tabularx}{\textwidth}{lXXXXXX}
\hline
Site & $x$ (\AA) & $y$ (\AA) & $z$ (\AA) & $D_x$ (\AA) & $D_y$ (\AA) & $D_z$ (\AA)\\
\hline
Pb1\hspace{2em} & 2.348348453254   &  -0.194416898148  &    0.000000000000   &   0.270838152536  &   -0.194998639113  &    0.000000000000  \\
Pb2 & 6.514591517139    &  3.905795745636   &   0.000000000000  &    0.280897054122  &   -0.250789561865   &   0.000000000000 \\
Pb3 & 10.303579037246   &   8.430702680087  &    0.000000000000  &   -0.086299588070   &   0.118113806051  &    0.000000000000 \\
Pb4 & 6.483787541004  &   -0.278698645131  &    0.000000000000   &   0.251256867342 &    -0.280443988453  &    0.000000000000 \\
Pb5 & 10.584724180095  &    3.886221429799   &   0.000000000000  &    0.196009344134  &   -0.271527480059    &  0.000000000000 \\
Pb6 & 14.428487837660 &     8.403032892886    &  0.000000000000   &  -0.116411160600    &  0.089280416493  &   -0.000000000000 \\
Pb7 & 2.332808528373  &   -0.298746489056  &    4.081492381624  &    0.255298227654  &   -0.299328230021  &    0.000000000000 \\
Pb8 & 6.469254068622  &    3.862452953432  &    4.081492381624    &  0.235559605605  &   -0.294132354068  &    0.000000000000 \\
Pb9 & 10.237223450768  &    8.464791293508  &    4.081492381624 &    -0.152655174548 &     0.152202419472  &    0.000000000000 \\
Pb10 & 6.527883079173  &   -0.233062403382  &    4.081492381624  &    0.295352405511 &     -0.234807746704 &      0.000000000000 \\
Pb11 & 10.686356778408 &     3.902916949018 &      4.081492381624 &      0.297641942447 &    -0.254831960839 &      0.000000000000 \\
Pb12 & 14.393032212768  &    8.465740937520  &     4.081492381624 &    -0.151866785492  &    0.151988461127 &     0.000000000000 \\
\hline
\end{tabularx}
\end{table}

\begin{table}[h]
\centering
\caption{DFT optimized Pb positions ($x, y, z$) and displacements ($D_x, D_y, D_z$) in the ``orthogonal'' domain structure of PbZrO$_3$. The displacement data were calculated by the difference between the relaxed structure and the highly symmetric (cubic) structure.}
\begin{tabularx}{\textwidth}{lXXXXXX}
\hline
Site\hspace{2em} & $x$ (\AA) & $y$ (\AA) & $z$ (\AA) & $D_x$ (\AA) & $D_y$ (\AA) & $D_z$ (\AA)\\
\hline
Pb1& 2.347710724285  &   -0.069232693074   &   0.000000000000 &     0.269503109617  &   -0.068703481215  &    0.000000000000 \\
Pb2& 6.517463570122  &    4.026211902598   &   0.000000000000 &     0.283899116596  &   -0.128615724401 &     0.000000000000 \\
Pb3& 10.320217811170  &    8.579687575475 &     0.000000000000 &    -0.068703481214 &     0.269503109617  &    0.000000000000 \\
Pb4& 14.415662406843 &    12.749440421313 &     0.000000000000 &    -0.128615724401 &     0.283899116596 &     0.000000000000 \\
Pb5& 6.495471131640  &   -0.150318265118 &     0.000000000000  &    0.260848516272 &    -0.148730750134 &     0.000000000000 \\
Pb6& 10.596367821280  &    4.015017026056 &     0.000000000000 &     0.206388367052  &   -0.138752297818 &     0.000000000000 \\
Pb7& 14.396605542953 &     8.569974679005 &     0.000000000000  &   -0.148730750134  &    0.260848516272 &     0.000000000000 \\
Pb8& 18.561940834126 &    12.670871368644 &     0.000000000000 &    -0.138752297818 &     0.206388367052  &    0.000000000000 \\
Pb9& 2.346557992189  &    -0.163085045946 &     4.081862045893 &     0.268350377521 &    -0.162555834087   &   0.000000000000 \\
Pb10& 6.478020218820  &    3.989425542861  &    4.081862045893 &     0.244455765294  &   -0.165402084138 &     0.000000000000 \\
Pb11& 10.226365458298 &     8.578534843379 &     4.081862045893 &    -0.162555834087 &     0.268350377521  &    0.000000000000 \\
Pb12& 14.378876047106  &   12.709997070010 &     4.081862045893   &  -0.165402084138  &    0.244455765294  &    0.000000000000 \\
Pb13& 6.531061124876   &  -0.111874260643 &     4.081862045893 &     0.296438509507 &    -0.110286745658  &    0.000000000000 \\
Pb14& 10.688001195157 &     4.019254966480 &     4.081862045893 &     0.298021740929 &    -0.134514357394  &    0.000000000000 \\
Pb15& 14.435049547428 &     8.605564672240 &     4.081862045893  &   -0.110286745658  &    0.296438509507 &     0.000000000000 \\
Pb16& 18.566178774551 &    12.762504742521 &     4.081862045893 &    -0.134514357394  &    0.298021740929  &    0.000000000000 \\
\hline
\end{tabularx}
\end{table}

\begin{table}[h]
\centering
\caption{DFT optimized Pb positions ($x, y, z$) and displacements ($D_x, D_y, D_z$) in the ``antiparallel'' domain structure of Pb(Zr$_{0.5}$Sn$_{0.5}$)O$_3$. The displacement data were calculated by the difference between the relaxed structure and the highly symmetric (cubic) structure.}
\begin{tabularx}{\textwidth}{lXXXXXX}
\hline
Site\hspace{2em} & $x$ (\AA) & $y$ (\AA) & $z$ (\AA) & $D_x$ (\AA) & $D_y$ (\AA) & $D_z$ (\AA)\\
\hline
Pb1& 2.311458739815   &  -0.161297982350  &   -0.014378185587  &    0.254406485328 &    -0.163770727184 &    -0.014231570742\\
Pb2& 6.419962099873 &     3.944376563194 &    -0.000786079640 &     0.243859481540 &    -0.177101095309  &   -0.000040420233\\
Pb3& 10.236463676407  &    8.348097120284 &    -0.005472552601 &    -0.058689305771 &     0.107614548111 &    -0.004127848632\\
Pb4& 6.350798830152  &   -0.236551454032  &   -0.000603664044  &    0.179642292821 &    -0.243969807568  &   -0.000163819515\\
Pb5& 10.453522975808 &     3.872738500175 &     0.012487616847 &     0.163316074632 &    -0.253684767031 &     0.013526505937\\
Pb6& 14.301733461072  &    8.303659641388 &     0.001216130683 &    -0.107523803949  &    0.058231460513  &    0.002854064334\\
Pb7& 2.256993054124 &    -0.246064484028 &     4.069557381831  &    0.200236879129   &  -0.248228923464 &     0.000110599090\\
Pb8& 6.352614576831   &   3.859106653444 &     4.078399090306 &     0.176808037990 &    -0.262062699661  &    0.009551352127\\
Pb9& 10.176364116344  &    8.348048160774 &     4.074258416161 &    -0.118492786342   &   0.107873894000 &     0.006009722543\\
Pb10& 6.433389075068 &    -0.170436809428 &     4.058361629357  &    0.262528617230   &  -0.177546857566  &   -0.010791923700\\
Pb11& 10.537079566134  &    3.925183773858 &     4.069690451956  &    0.247168744451 &    -0.200931187949 &     0.001135943460\\
Pb12& 14.301853212746  &    8.362085371207  &    4.059687235761 &    -0.107107972783  &    0.116965495730 &    -0.008268228174 \\
\hline
\end{tabularx}
\end{table}

\begin{table}[h]
\centering
\caption{DFT optimized Pb positions ($x, y, z$) and displacements ($D_x, D_y, D_z$) in the ``orthogonal'' domain structure of Pb(Zr$_{0.5}$Sn$_{0.5}$)O$_3$. The displacement data were calculated by the difference between the relaxed structure and the highly symmetric (cubic) structure.}
\begin{tabularx}{\textwidth}{lXXXXXX}
\hline
Site\hspace{2em} & $x$ (\AA) & $y$ (\AA) & $z$ (\AA) & $D_x$ (\AA) & $D_y$ (\AA) & $D_z$ (\AA)\\
\hline
Pb1& 2.309761187667  &   -0.039885067148  &    8.130140155805 &     0.253233329913 &    -0.041647886548  &   -0.009198320990\\
Pb2& 6.422979111757  &    4.059370430516 &     8.136567938535 &     0.249381244137  &   -0.059462398749   &  -0.002662266171\\
Pb3& 10.249019990937 &     8.489136169044  &    8.129923611628  &   -0.041647886548 &     0.253233329914   &  -0.009198320990\\
Pb4& 14.348275488601  &   12.602354093134 &     8.136351394359   &  -0.059462398749   &   0.249381244137   &  -0.002662266171\\
Pb5& 6.359739961554   &  -0.112432700972  &    0.002416030259 &     0.189790147448  &   -0.118087745216   &   0.002497234323\\
Pb6& 10.462554656156  &    4.002107631775    &  0.008720537233   &   0.175534832185   &  -0.120617422334  &    0.008910013385\\
Pb7& 14.286002088620  &    8.429585211423 &     0.002199486082  &   -0.118087745216    &  0.189790147449    &  0.002497234323\\
Pb8& 18.400542421367   &  12.532399906025    &  0.008503993057   &  -0.120617422334  &    0.175534832185  &      0.008910013385\\
Pb9& 2.271610898165   &  -0.115044505295    &  4.071965149737   &   0.215021962583   &  -0.116868402523   &   0.002309445350\\
Pb10& 6.365793573492   &   3.981767545096 &     4.079894654584   &   0.192134628046   &  -0.137126361997  &    0.010347222286\\
Pb11& 10.173860552790  &    8.450985879542 &     4.071748605561 &    -0.116868402523   &   0.215021962583    &  0.002309445350\\
Pb12& 14.270672603181  &   12.545168554869  &    4.079678110408 &    -0.137126361997   &   0.192134628046   &   0.010347222286\\
Pb13& 6.437708068474  &   -0.052768885629  &    4.058545973018  &    0.267819332196   &  -0.058362852045  &   -0.011055595326\\
Pb14& 10.536091534816 &     4.037903450890 &     4.066754913299  &    0.249132788673 &    -0.084760525391    & -0.002738382957\\
Pb15& 14.345665903963  &    8.507553318344  &    4.058329428842  &   -0.058362852045   &   0.267819332196  &   -0.011055595326\\
Pb16& 18.436338240483 &    12.605936784686  &    4.066538369123    & -0.084760525391  &    0.249132788673 &    -0.002738382957\\
\hline
\end{tabularx}
\end{table}